\documentclass[conference]{IEEEtran}
\IEEEoverridecommandlockouts

\usepackage{cite}
\usepackage{amsmath,amssymb,amsfonts}
\usepackage{algorithmic}
\usepackage{hyperref}
\usepackage{diagbox}

\usepackage{graphicx}
\usepackage{textcomp}
\usepackage{xcolor}
\usepackage{graphicx}
\usepackage{subfigure}
\usepackage[linesnumbered,ruled,vlined]{algorithm2e}

\def\BibTeX{{\rm B\kern-.05em{\sc i\kern-.025em b}\kern-.08em
    T\kern-.1667em\lower.7ex\hbox{E}\kern-.125emX}}

\begin{document}
\IEEEoverridecommandlockouts

\IEEEpubid{\makebox[\columnwidth]{978-0-7381-1151-3/20/\$31.00 \copyright 2020 IEEE \hfill} \hspace{\columnsep}\makebox[\columnwidth]{ }}


        

\title{Lossy Medical Image Compression using Residual Learning-based Dual Autoencoder Model\\
%
}


\author{\IEEEauthorblockN{Dipti Mishra\IEEEauthorrefmark{1},
Satish Kumar Singh\IEEEauthorrefmark{2}, and Rajat Kumar Singh\IEEEauthorrefmark{3}}

\IEEEauthorblockA{%
    \IEEEauthorrefmark{1}Department of Electronics \& Communication Engineering \\
  }
  \IEEEauthorblockA{%
    \IEEEauthorrefmark{2}Department of Information Technology \\
  }
  \IEEEauthorblockA{%
    \IEEEauthorrefmark{3}Department of Electronics \& Communication Engineering \\
  }
Indian Institute of Information Technology Allahabad, Prayagraj, 211015, India\\

Email: \IEEEauthorrefmark{1}dipti.mishra28@gmail.com,
\IEEEauthorrefmark{2}sk.singh@iiita.ac.in,
\IEEEauthorrefmark{3}rajatsingh@iiita.ac.in}

\maketitle
\begin{abstract}
In this work, we propose a two-stage autoencoder based compressor-decompressor framework for compressing malaria RBC cell image patches. We know that the medical images used for disease diagnosis are around multiple gigabytes size, which is quite huge. The proposed residual-based dual autoencoder network is trained to extract the unique features which are then used to reconstruct the original image through the decompressor module. The two latent space representations (first for the original image and second for the residual image) are used to rebuild the final original image. Color-SSIM has been exclusively used to check the quality of the chrominance part of the cell images after decompression. The empirical results indicate that the proposed work outperformed other neural network related compression technique for medical images by approximately 35\%, 10\% and 5\% in PSNR, Color SSIM and MS-SSIM respectively. The algorithm exhibits a significant improvement in bit savings of 76\%, 78\%, 75\% \& 74\% over JPEG-LS, JP2K-LM, CALIC and recent neural network approach respectively, making it a good compression-decompression technique.
\end{abstract}

\begin{figure*}[htbp]
\centerline{\includegraphics[width=0.99\textwidth]{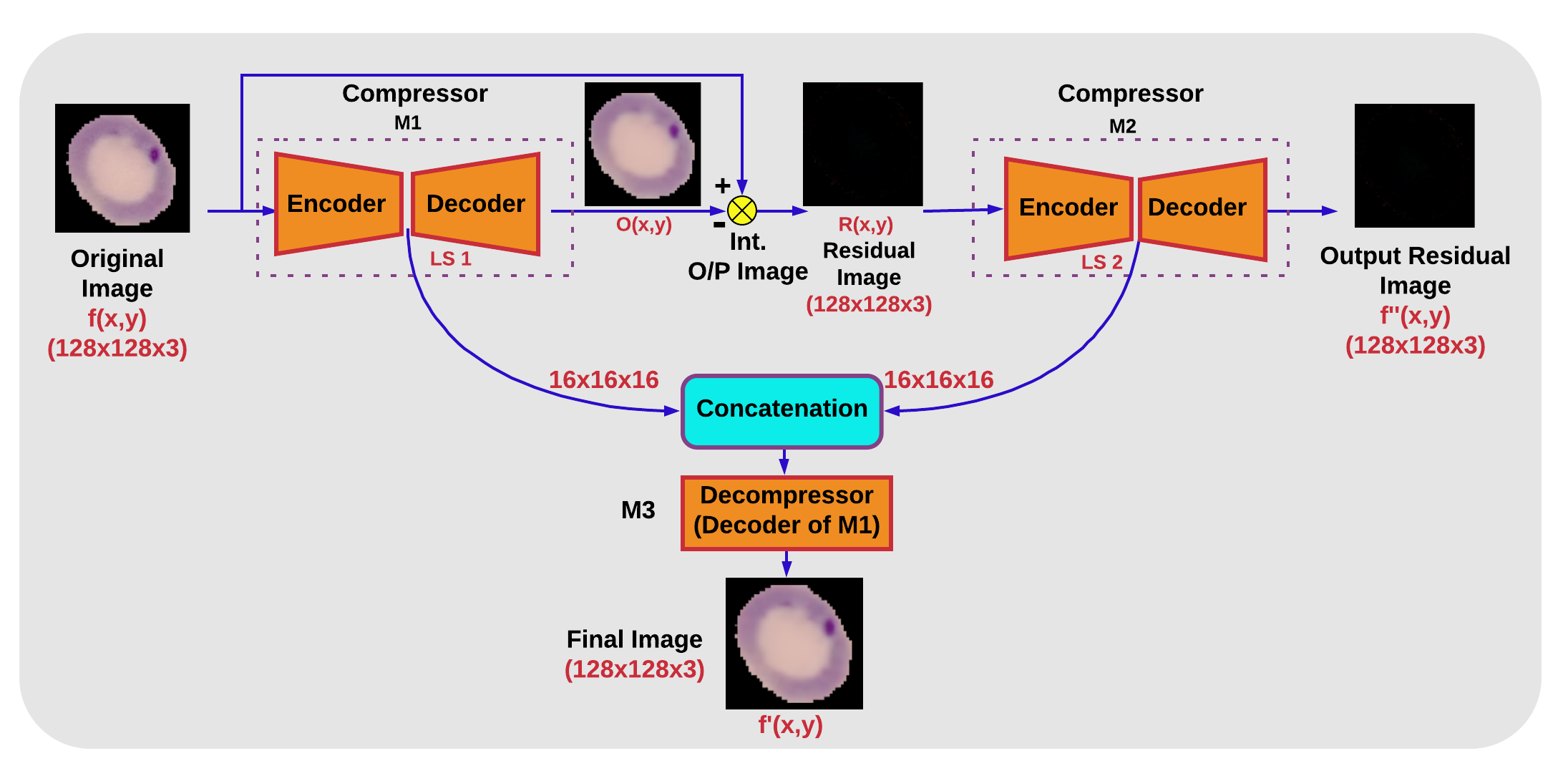}}
\caption{Block diagram of the proposed compression-decompression model.}
\label{fig:block13}
\end{figure*}
\section{Introduction}
\IEEEPARstart{D}{eep} learning techniques are now mostly used for diverse applications in image classification, object detection, and segmentation task. Rapid development in this field is made possible with the help of fast computing GPU resources. With the advancement in resources, image compression researchers are also diving and digging for these deep architectures available to date to accomplish the task of image compression~\cite{fullresolution,semantic,optimized,lossy}. Applying the principle of exploiting domain knowledge of image processing with the deep neural network architectures, proved to be better for image compression over conventional pre-deep learning era algorithms. The traditional image compression algorithms were dependent on transform and quantization analysis, and among them, mostly are lossy in nature. Mostly the researchers are confined to lossy image compression due to various constraints of quality distortion and limitations on compression ratio. In the medical field, pathologists used to diagnose diseases by analyzing large images using a recently developed imaging technique called Whole Slide Imaging (WSI)~\cite{wsi1,wsi2}. WSI is a process to scan the glass slides to produce a digital slide. These images provide higher resolution and features in multi layers to extract very minute and fine details. The size of these images ranges from many gigabytes, which need a lot of storage and transmission bandwidth. 
So, there is an alarming need for highly efficient lossless image compression algorithms to compress information-rich images like medical images.
For compressing these images, lossy compression for these whole slide images can affect the diagnosis process due to the information loss. Therefore, lossless image compression is the optimal approach to preserve relevant information in the images where any type of data loss is not acceptable; for example, medical images, satellite images, or images to be sent to army war field areas. 
JPEG 2000~\cite{jpeg2000}, being lossless and lossy both, is a region of interest (ROI) based compression scheme involving multi-band decomposition through discrete wavelet transform (DWT). ROI implies that the different parts of the image can be encoded with variable bit rates providing variable image qualities. 
The uniqueness of this compression scheme is its flexibility of scalable or progressive decoding in resolution and quality based on the availability of the receiver system data rate (related to different compression ratios based on different bit-rate suitability). Due to this property, the encoder and decoder employed in its architecture are quite complicated and costly.
Specifically, for lossless compression, the best available technique in literature is JPEG in lossless mode (JPEG-LS)~\cite{jpegls}, a prediction based effective approach based on a median edge detector (MED) with an additional entropy coding technique namely Golomb Rice Encoding (GRC)~\cite{grc}. Context adaptive lossless image compression (CALIC)~\cite{calic} was also heavily employed in literature based on the gradient adjusted predictor (GAP). 
It is based on MANIAC (Meta-Adaptive Near-zero Integer Arithmetic Coding), a context-adaptive method based on growing decision trees. Edge directed predictor (EDP)~\cite{edp} was another approach which is optimized with least square techniques. 
In 2014, Hesabi et al.~\cite{memory} exploited principle component analysis (PCA)~\cite{pca} to learn the relationship between input samples and compressed representation to learn cross-image correlation. In~\cite{memory}, PCA was exploited to reduce the inter and intra image redundancies. The method outperformed JP2K-LM (JPEG 2000 lossless mode)~\cite{jpeg2000}, JPEG-LS~\cite{jpeg}, CALIC, and bzip2\footnote{\url{http://www.bzip.org/}} algorithm. However, the algorithm was only suitable for the correlated medical image dataset and not able to encounter sharp images, i.e., infected and non-infected cell images. 
On the other hand, the essential features extracted from the convolutional neural network (CNN) help to decide the salient regions so that context-specific image compression can be achieved by allocating variable bits to different parts of the region. The salient region can be identified with the help of essential features extracted through CNN. This unique property of feature extraction forced the researchers to shift their domain to deep learning algorithms. 
All the compression techniques discussed so far have a compression ratio limited to 2:1 only.
Then in 2016, Shen et al.~\cite{shen} demonstrated a region of interest (ROI) based lossless cum lossy compression scheme for medical images. The approach exploited very less dissimilarity property between adjacent ROIs. The binary ROI feature maps are encoded by arithmetic coding~\cite{arith}. Later on, Shen et al.~\cite{shen2} provided a lossless compression scheme followed by the Golomb Rice encoding technique (GRC)~\cite{grc} on malaria-infected cells. However, the results on infected cells are not as good as in non-infected cells, showing that deep learning architectures have become less useful for infected cell image compression. 

Here, we focus on CNN based image compression for medical images like RBC blood cells. We have developed an algorithm based on residual error calculation for ROI based lossy compression. The architecture is simply a CNN designed for training of both correlated and uncorrelated type of cell images. In this work, we propose a de-correlation based two-stage autoencoder network designed for lossy image compression, whose functionality is to extract features to reduce inter-image redundancies in addition to intra-image de-correlation. The proposed residual-based dual autoencoder is used to extract the significant features of non-infected and malaria-infected RBC cells. We have exploited the principle of deep residual autoencoding based on Restricted Boltzmann Machine (RBM) mentioned in~\cite{red,greedy}.
When implemented and assessed with performance metrics, the network worked well for both infected and non-infected cells, however much better for compressing infected cells with no loss in ROI. The key significant contributions are summarized as follows:
\begin{itemize}
\item First autoencoder is utilized to learn low-frequency components; however, the second autoencoder learns the high-frequency components of the residual image obtained. The second learning thereby helps to retain the high-frequency components like edges and textures etc., and hence helped to produce good quality images and calling it a lossy compression scheme.
\item The algorithm has incorporated ROI based encoding to ensure variable bit coding dependent on the spatial content of the image.
\item Since the medical images are of various colors and color symbolizes essential information, so we have directly incorporated color based structural similarity (C-SSIM) into account without exploiting color to gray and gray to color conversion in the pre-processing to avoid information loss. So, the use of color-SSIM ensures accurate subjective assessment.
\end{itemize}

The rest of the paper details is organized as follows. 
Section~\ref{sec:proposed} discusses the proposed compression-decompression framework based on dual autoencoders. Section~\ref{sec:dataset} illustrates the dataset and other experimental set up details. Section~\ref{sec:results} throws a light on the empirical results obtained with the proposed algorithm. Section~\ref{sec:conclusion} ends with a conclusion of the work described in this paper.

\section{Proposed Dual Autoencoder based Compression Framework}
\label{sec:proposed}
We have designed a codec network by cascading the compressor and decompressor module, exploiting a deep autoencoder. The residual dual deep architecture technique is used to achieve lossy image compression on Malaria infected cell images. The network architecture for the proposed algorithm is shown in Fig.~\ref{fig:block13}. The compressor module (M1) consists of an encoder-decoder architecture. Table~\ref{details} shows the layer-wise network details for the compression ratio of $6:1$. Fully connected layer has not been used in the proposed encoder-decoder network making it a fully convolutional network (FCN) in nature.
Initially, the deep autoencoder (M1) has been trained to get a compressed latent space representation (LS 1) of $16\times16\times16$ size, which is then reconstructed by the decoder to obtain intermediate output image $O(x,y)$ of $128\times128\times3$ size. We further experimented for improving the quality of the obtained image. For that, the residual image $R(x,y)$ is calculated by subtracting this intermediate output $O(x,y)$ from the original image $f(x,y)$. A factor of 128 then scales this residual image because of very little pixel intensity. The scaled residual is then substantially applied as the input to the same autoencoder architecture (M2) for residual learning only. This is to obtain a mapping of latent space representation with the residuals to get high-frequency information (edges, textures or boundary). The final image is then reconstructed through this model with the help of compressed latent space representation (LS 1 and LS 2). These latent space representations LS 1 and LS 2 are concatenated followed by upscaling with the help of convolution and pooling layers. The decompressor module (M3) or decoder of module M1 or M2 acts by taking this concatenated and upscaled representation to decode them for the final reconstruction of the image. 
The visual quality of the reconstructed image inferred that the dual autoencoder model performed well in improving the quality of images in terms of these compression performance metrics. 

\begin{table}[htb]
\caption{Architecture details of the proposed Compressor (M1/M2) and Decompressor module (M3) for $6:1$ compression ratio.}
\begin{center}
\begin{tabular}{p{4cm} | p{3cm} }
\hline
\hline
Encoder & Decoder\\ 
\hline
$3\times3$, 64 Conv, strides $2\times2$, ReLU & Dense, 16, ReLU \\
$2\times2$ Maxpooling $\downarrow$ &   $3\times3$, 32 Conv, ReLU            \\
$3\times3$, 32 Conv, strides $2\times2$, ReLU & $2\times2$ Upsampling  $\uparrow$  \\ 
$2\times2$ Maxpooling  $\downarrow$   &   $3\times3$, 32 Conv, ReLU                     \\
$3\times3$, 32 Conv, strides $2\times2$, ReLU &   $2\times2$ Upsampling $\uparrow$ \\ 
$2\times2$ Maxpooling  $\downarrow$ &     $3\times3$, 64 Conv, ReLU                 \\
Dense, 16, ReLU &   $2\times2$ Upsampling $\uparrow$ \\ 
  & $3\times3$, 3 Conv, Sigmoid \\
\hline
\hline
\end{tabular}
\label{details}
\end{center}
\vspace{-5mm}
\end{table}

\begin{table*}
\caption{Comparison of proposed algorithm with the related deep learning based method.}
\begin{center}
\begin{tabular}{c|c |c|c| c}
\hline
\hline
Algorithm & PSNR~(dB)/C-SSIM/MS-SSIM & Encoding Time & Decoding Time & Memory Used \\
\hline
Shen\textquotesingle s~\cite{shen2} & 27.50/0.9423/0.9412   & 15ms & 15ms & 11.59 GB  \\
Proposed (with M1) (6:1) & 29.33/0.9693/0.9634 & 5ms & 5ms & 9.11 GB \\
Proposed (with M1, M2 \& M3) (6:1)  & 35.91/0.9878/0.9704  & 6.67ms & 10ms & 11.23 GB \\
\hline
\hline
\end{tabular}
\label{resultvalues}
 \vspace{-4mm}
\end{center}
\end{table*}

\section{Dataset and Experimental Settings}
\label{sec:dataset}
The network has been trained on the Malaria Cell Images Dataset\footnote{\url{https://www.kaggle.com/iarunava/cell-images-for-detecting-malaria}}, which consists of 27,558 cell images having equal samples of parasitic and non-infected cells. It is used to train the proposed network to capture the discriminative features of both type of cell images. The parasitic cell has a purple ring formation, which differentiates it from non-infected cells. 
In this dataset, the image size varies from $130\times130\times3$ to $360\times360\times3$, so we have resized the complete dataset to $128\times128\times3$. Accordingly, 22,222 and 5334 images are used for training and testing the network respectively. 
To achieve accurate decision making, we have applied data cleaning/whitening for getting reliable datasets. Data whitening is used to make the data less redundant so that all the features become uncorrelated with the same variance. By this way, correlation is reduced by merely projecting the dataset into eigen vectors, which is then followed by normalization.

 The dual-stage autoencoder network has been trained with mean square loss (MSE) function~\cite{mse}. To achieve global minima, the network has been trained for 100 epochs with a batch size of 8.
 We have used the Adam optimizer with learning rate as $0.001$. Moreover, the momentum parameters $\beta_1=0.9$, $\beta_2=0.999$ are selected which are reported to produce better results in~\cite{mse,dipti}. 
 
 The proposed algorithm has been evaluated for the quality of reconstructed images with the help of peak signal-to-noise ratio (PSNR)\cite{psnr} and multi-structural similarity (MS-SSIM)\cite{msssim}. 
 One more thing to attract the attention of the researchers is that for these colored medical cell images, Color-SSIM (C-SSIM)~\cite{colorssim,colorssim2} has been exploited to assess the color quality of the images.
 It is because color images reflect more meaningful information than the corresponding gray-scale images.
 For color-SSIM, the image is modeled as a combination of four functions i.e., luminance comparison, contrast comparison, structural comparison, and the color comparison. By definition, $Color-SSIM=l(x,y).C(x,y).S(x,y).C_{r}(x,y)$
where, $l(x,y)$, $C(x,y)$, $S(x,y)$ and $C_{r}(x,y)$ are the luminance, contrast, structure and color comparison factor respectively between the original and reconstructed distorted images. $C_{r}(x,y)$ is defined as
$C_{r}=1-\frac{1}{k}\bigtriangleup E(x,y)$, where $\bigtriangleup E(x,y)$ is color fidelity value which is calculated between the S-CIELAB conversions of the two images $X$ and $Y$; $k$ is a weighting constant,
$\bigtriangleup E=\sqrt{(L_{1}^{*}-L_{2}^{*})^2+(a_{1}^{*}-a_{2}^{*})^2+ (b_{1}^{*}-b_{2}^{*})^2)}$.
Here, $L^{*}$, $a^{*}$ and $b^{*}$, represents the values from darkness to lightness (in the intensity range from 0-100), greenness to redness (in the intensity range from -128 to +127) and blueness to yellowness (in the intensity range from -128 to +127) respectively. Accordingly $L_{1}^{*}$, $a_{1}^{*}$, $b_{1}^{*}$ and $L_{2}^{*}$, $a_{2}^{*}$, $b_{2}^{*}$ are defined for the two images $X$ and $Y$ respectively.
We have used Python, an open-source coding platform with Keras and Tensorflow, deep learning framework for all the experiments. The network is trained on 16 GB CPU RAM, 12 GB GPU enabled system TITAN X (Pascal)/PCIe/SSE2.

\section{Results \& Discussion}
\label{sec:results}
\begin{figure}[htb]
\centering
{\includegraphics[width=\linewidth,height=8cm]{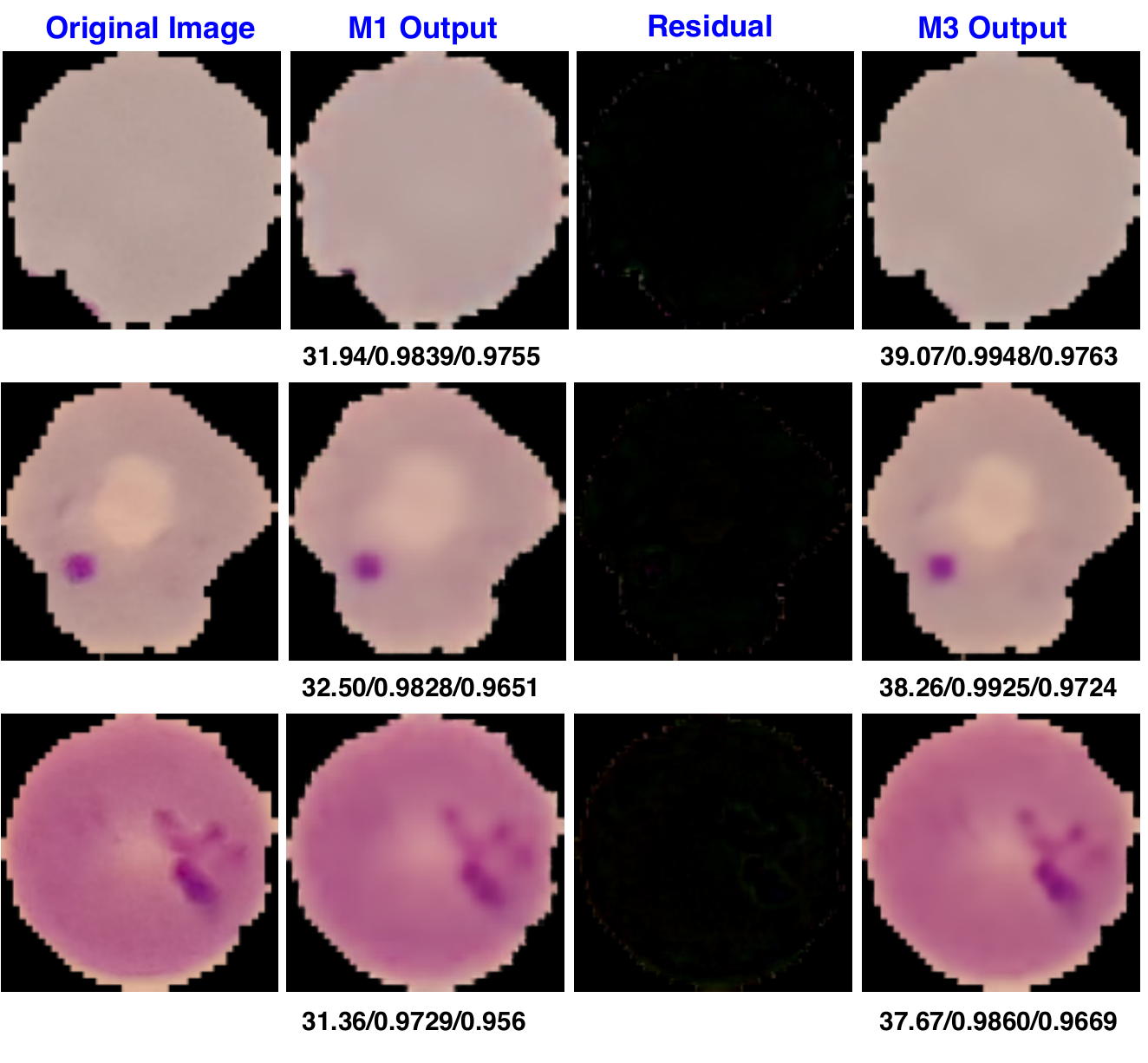}}
\caption{Visual quality comparison (PSNR (dB)/C-SSIM/MS-SSIM) of final reconstructed images (M3 output) with the original images and intermediate output images (M1 output).}
\label{fig:images1}
\vspace{-4mm}
\end{figure}

The proposed network containing a two-stage autoencoder network based on residual learning helped to preserve the fine details in infected cell images. Here, we found that the storage space required to store the latent space representation is 12 times less than the original image size, making the compression ratio to $6:1$. Since for training, the first autoencoder (M1) is provided with the raw images and the second autoencoder (M2) is provided with the residual images, it can be inferred that M1 and M2 are learning the low and high-frequency components respectively. The additional implementation of the second (M3), which is a decompressor module, further improved the quality of reconstructed malaria cell images as the final output is better than $O(x,y)$ in terms of performance measures as shown in Table~\ref{resultvalues}. The empirical results indicate that the proposed work outperformed Shen\textquotesingle s compression technique for medical images by approximately 35\%, 10\% and 5\% in PSNR, Color SSIM and MS-SSIM respectively. Also, the output images are found to contain less number of blocking and ringing effects as can be seen from images shown in Fig~\ref{fig:images1}. It is because the network is able to retain both low-frequency and high-frequency components which are learned in two-phases. We have compared the original image with intermediate output and final output separately. As color images reflect more meaningful information than the corresponding gray-scale images, color-SSIM has been used for proper assessment. 
 The first example, shown in Fig.~\ref{fig:images1} being a normal cell image, shows a high value of MS-SSIM, as the algorithm need not focus on any region of interest, and each region is compressed uniformly. But for the other two images, more bits are allocated for ROI, and less on other regions, these enhanced values of the metrics for ROI claims to be another contribution. 
For the sake of comparison, we have reproduced the results of Shen\textquotesingle s algorithm. The bit-rate comparison (at CR=6:1) with the state-of-the-art method reveals that the proposed algorithm is able to perform excellently at an extremely low bit-rate (1.4 bpp) for providing lossy compression as shown in Table~\ref{bpp}. The algorithm exhibits a significant improvement in bit savings of 76\%, 78\%, 75\% \& 74\% over JPEG-LS, JP2K-LM, CALIC and recent neural network approach respectively. The algorithm design is universal for all types of images, but we develop a lossy compression algorithm to generate distortion-less images at extremely low bit-rates (bpp) and high compression ratio (much beyond 2:1).
\begin{table}
\caption{Compression gain (bpp) comparison with benchmark and state-of-the-art algorithms.}
\begin{center}
\begin{tabular}{p{1cm}|p{1cm}| p{1cm}|p{1.4cm} |p{1.2cm}|p{1cm} }
\hline
\hline
Method & JPEG-LS~\cite{jpegls} & JP2K-LM~\cite{jpeg2000} & CALIC~\cite{calic} & Shen\textquotesingle s~\cite{shen2} & Proposed \\
\hline
Bit-rate (bpp) & 5.83 & 6.34 & 5.52 & 5.34 & 1.40 \\
\hline
\hline
\end{tabular}
\label{bpp}
\end{center}
 \vspace{-6mm}
\end{table}
It is found that the results obtained are far better than the mentioned approaches due to residual learning technique which is responsible for preserving finer details. The training time undertaken by M1 and M2 modules is 75s and 25s per epoch, respectively. So, the total time taken by the network to train is approximately 2 hours and 70 minutes respectively. Also, the average residual generation time is 79ms, which is very less. The testing time to decode one image is around 10ms, which is 1.5 times the encoding time.
The comparison on the basis of compression parameters like encoding time, decoding time and memory resource utilization is given in Table~\ref{resultvalues}. The testing time also indicates that the proposed network is not too much complex and can be practically realized.
Additionally, we have also applied the Huffman encoding and decoding technique on the latent space representations (to obtain bit-stream), further to reduce the file size to be stored or transmitted. However, when implemented, it did not contributed in further reducing file size. This is due to the minimal size of the training images used since Huffman coding is very efficient in encoding large-sized images. We have also tried to incorporate context adaptive binary adaptive coding (CABAC) as an entropy coding on the residual obtained, but it didn\textquotesingle t provide much improvement.
As far as we know, this is one of the good lossy image compression approaches for medical images giving better results than the mentioned approach as high-frequency components are learned separately through residual learning.

\section{Conclusion}
\label{sec:conclusion}
We have implemented the dual autoencoder based compression-decompression algorithm using deep learning technique on malaria cell images. The dual autoencoder model is utilized to exploit both low and high frequency details of the image. It is clear that, the fine details in the ROI are not affected after decompression, which does not affect the diagnosis by any medical expert. The proposed framework reports to be a good ROI based lossy compression scheme, producing good quality images with minimum information loss. The algorithm successfully outperformed JPEG-LS, JP2K-LM, CALIC and other deep learning based state-of-the-art approaches. From the experiments, it is also concluded that the compression of the image not only depends upon the model architecture but also on the type and size of the images on which the model is being trained. For achieving more improvement with Huffman coding, large size images should be chosen for training with comparatively deep architecture, which can be a future scope of the work.

\bibliographystyle{unsrt}

\bibliography{references}
\end{document}